\documentclass[aps,twocolumn]{revtex4-1}
\usepackage{amsfonts}
\usepackage{amsmath}
\usepackage{amssymb}
\usepackage{bm}
\usepackage[pdftex]{graphicx}
\begin{document}
\title{Fluctuating pancake vortices revealed by dissipation of
Josephson vortex lattice}
\author{A. E. Koshelev}
\affiliation{Materials Science Division, Argonne National
Laboratory, Argonne, Illinois, 60439, USA}
\author{A. I. Buzdin}
\affiliation{Institut Universit\'{e} de France and Universit\'{e} de
Bordeaux, LOMA, UMR 5798, F-33405 Talence, France}
\author{I. Kakeya}
\affiliation{Department of Electronic Science and Engineering, Kyoto
University, Nishikyo, Kyoto 615-8510 Japan}
\author{T. Yamamoto}
\affiliation{Institute of Materials Science, University of Tsukuba,
Tsukuba, Ibaraki 305-8573 Japan}
\author{K. Kadowaki}
\affiliation{Institute of Materials Science, University of Tsukuba,
Tsukuba, Ibaraki 305-8573 Japan}

\date{\today }

\begin{abstract}

In strongly anisotropic layered superconductors in tilted magnetic
fields the Josephson vortex lattice coexists with the lattice of
pancake vortices. Due to the interaction between them, the
dissipation of the Josephson-vortex lattice occurs to be very
sensitive to the presence of the pancake vortices. If the c-axis
magnetic field is smaller then the corresponding lower critical
field, the pancake stacks are not formed but the individual pancakes
may exist in the fluctuational regime either near surface in
large-size samples or in the central region for small-size mesas. We
calculate the contribution of such fluctuating pancake vortices to
the c-axis conductivity of the Josephson vortex lattice and compare
the theoretical results with measurements on small mesas fabricated
out of Bi$_{2}$Sr$_{2}$CaCu$_{2}$O$_{8+\delta}$ crystals. A
fingerprint of fluctuating pancakes is characteristic exponential
dependence of the c-axis conductivity observed experimentally. Our
results provide strong evidence of the existence of the fluctuating
pancakes and their influence on the Josephson-vortex-lattice
dissipation.

\end{abstract}
\maketitle

\section{Introduction}

Vortex physics in layered superconductors is extremely rich and
interesting. This is mostly related to the fact that the magnetic
field parallel to the layers penetrates in the form of the Josephson
vortices (JVs) \cite{BulClem91} while the perpendicular field
creates the stacks of the pancake vortices (PVs) \cite{PVgen}. In
the tilted magnetic field the attractive interaction between JVs and
PVs \cite{Koshelev99} leads to the creation of many unusual vortex
states: mixed chain-lattice states \cite{chain-lattice}, JVs
decorated by PVs \cite{DecoratedJV}, etc (see as a review Ref.\
\onlinecite{Bending05}). The interaction between the PV and JV
lattices is revealed also in their dynamic properties. Dynamics of
the JV lattice in Bi$_{2}$Sr$_{2}$CaCu$_{2}$O$_{8+\delta}$ (Bi2212)
mesas have been explored by several experimental groups
\cite{Lee,Hechtfischer,Latyshev,OoiPRL02,RectSmallStacks,KakeyaPRB:2009}. Due to the
weak intrinsic dissipation, the friction force acting on the moving
JV lattice is small. That is why even the presence of a small amount
of the PV stacks strongly affects the dynamic of the JVs.
The physical reason is that the moving JVs induce displacements of
the PVs which have large viscous friction due to the normal cores.
This strongly enhances the JV friction leading to decrease of the
measured c-axis resistivity of the sample.
Recently, this effects has been studied in details in Refs.\
\onlinecite{Koshelevetal06,LeePRB10}. To our knowledge, this is the
only known situation in which adding vortices to a superconductor
reduces its resistivity.  Therefore, studying the dynamic properties
of the JV lattice, we may obtain the information about the presence
of the PVs.

In the present paper we use this tool to study the penetration of
the PVs in superconducting mesas of the Bi2212 intrinsic Josephson
junctions. If the magnetic field is applied exactly along the
layers, then the c-axis resistivity is governed by the friction
force acting on the moving JV lattice. When the perpendicular
component of magnetic field $H_z$ is switched on, the PV stack can
enter into the sample when $H_{z}$ reaches the value of the
perpendicular low critical field $H_{c1}^{c}$.\cite{demag-footnote}

The c-axis field penetration is rather special in the case of
layered superconductors with weak Josephson coupling between the
layers. Indeed, the field starts to penetrate in the form of the PVs
created near the surface in the fluctuational regime. The energy of
a single PV near the surface has been calculated in Ref.\
\onlinecite{BuzFeinb92} (see also \cite{MintsSBPRB96}) and above
certain field (somewhat lower than $H_{c1}^{c})$ it has a minimum at
distance of the order of London penetration depth $\lambda$ from the
surface.
In small-size samples with lateral dimensions comparable with the
London penetration depth this energy minimum appears in the center
of the sample.
The concentration of the fluctuation PVs is determined by this
minimum energy, which, in turn, depends on the perpendicular
magnetic field.

The presence of these fluctuational PVs increases the friction of
the JVs and then it may be directly monitored by the measurement of
the c-axis resistivity. We obtain a very characteristic exponential
dependence of the c-axis conductivity vs $H_{z}$, reflecting the
varying density of the thermally activated PVs near the surface
or in the center of small samples. This dependence is clearly
observed experimentally.
These results demonstrate that the JVs serve as a perfect tool to
detect the presence of the PVs in the fluctuational regime.

The behavior of the JV lattice has special features in the mesas
with lateral sizes of few Josephson lengths used in our experiments.
In this case bulk interactions favoring the triangular lattice
compete with edge interactions favoring the rectangular lattice. Due
to this competition, the JV lattice undergoes a series of structural
phase transitions between the triangular and rectangular
configurations.\cite{MachidaPRL06,MagOscTheory}  The regions of
rectangular lattice dominate at large fields. As in the rectangular
lattice the in-plane current are absent, its interaction with PVs is
very weak. Therefore the fluctuational PVs have strong influence
only on dissipation of the triangular lattice. Due to
commensurability effects, the c-axis resistivity has pronounced
oscillations as function of the magnetic field.\cite{OoiPRL02} The
period of these oscillations changes from half flux quantum per
junction at small fields to one flux quantum per junction at large
fields \cite{RectSmallStacks,KakeyaPRB:2009} and the crossover field
between these regimes is proportional to mesa
width.\cite{MagOscTheory}

The article is organized as follows. In section \ref{Sec:Theory} we
describe theory of the effect. In particular, in subsection
\ref{Sec:FluctPanc} we derive the general formalism describing the
response of the statistical ensemble of PVs in the static potential
(creating by interaction with the surface) to the moving periodic
potential (due to the interaction with moving JV lattice). We
provide the expression for the resulting dissipation. In subsection
\ref{Sec:PVEnergy} we detail the potential of the PV interaction
with surface in presence of the perpendicular field $H_{z}$.
We also consider potential energy of a PV in finite-size samples
with simple geometries (strip and cylinder).
The energy of the interaction between PV located near the surface
and the dense lattice of the JV's is calculated in subsection
\ref{Sec:PV-JV}. We analyze the general expression for the PV
contribution to the JV lattice conductivity in section
\ref{Sec:JVConduct}. In section \ref{Sec:Exper} we present the experimental data on the c-axis conductivity of Bi$_{2}$Sr$_{2}$CaCu$_{2}$O$_{8+\delta}$ mesas with
different lateral sizes in tilted magnetic field and compare the results with
theoretical predictions.
\begin{figure}[ptb]
\begin{center}
\includegraphics[width=3.4in]{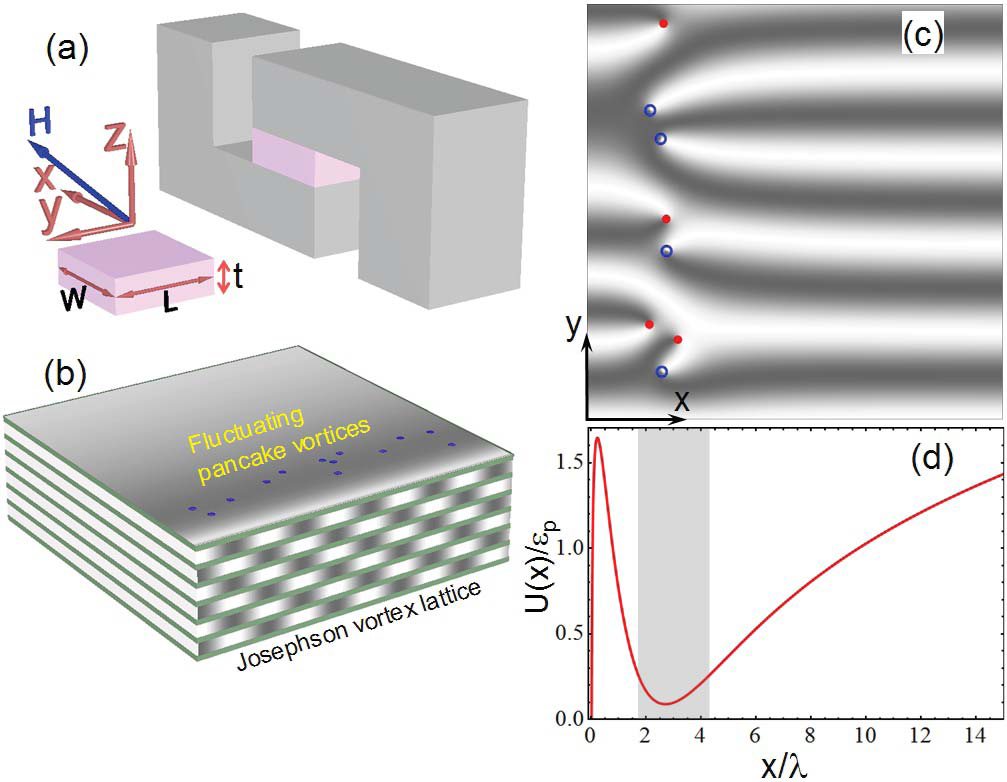}
\end{center}
\caption{(color online) These schematic figures illustrate mesa
geometry and the interaction between the dense JV lattice and
fluctuational PVs near the surface. (a) Right part shows outline of
the structure fabricated from Bi2212 crystal to study c-axis
transport through highlighted mesa region. Left part shows
definitions of axes and mesa sizes. (b) Three-dimensional sketch of
the JV lattice inside the junctions (front face) and the fluctuating
PVs near the surface (top face). (c) The graylevel plot of the
cosine of phase difference between two neighboring layers which have
periodic modulation due to the JV lattice and perturbations due to
randomly located PVs in two layers shown by open and closed circles.
(d) The energy profile of PVs near the surface in the c-axis
magnetic field close to the lower-critical field. The PV density is
enhanced near the potential minimum. This highlighted region near
the minimum of the potential gives dominating contribution to
dissipation of the JV lattice. } \label{Fig-JVPVSurfSchem}
\end{figure}


\section{Theory \label{Sec:Theory}}

\subsection{Fluctuation pancake vortices in static and moving periodic
potentials: General consideration \label{Sec:FluctPanc}}

We consider the geometry with the parallel-to-layers magnetic field
directed along x-axis $H_{\parallel}=H_{x}$ and therefore, the
perpendicular current (along c-axis ) will drive the JV lattice
along y-axis, see Fig.\ \ref{Fig-JVPVSurfSchem}(a). The PV located
at distance $x$ will interact with the surface ($yz$ plane) and with
a current screening the perpendicular field $H_{z}$ as well as with
a moving JV lattice. The first two contributions are described by
the static potential $U_{0}(x)$. The moving periodic potentials of
the JV lattice is $\tilde{U} (x)\cos (ky-\omega t)$, corresponding
to the velocity $v=\omega/k$. This motion generates an electric
field $E_{z}=H_{x}v/c.$ Therefore, we are interested in the behavior
of the PVs in the two-dimensional potential \bigskip
\begin{equation}
U(\mathbf{r},t)=U_{0}(x)+\tilde{U}(x)\cos(ky-\omega
t).\label{Potential}
\end{equation}
We start from a general consideration of a model in which the PVs
are considered as an ensemble of particles with concentration $n$ in
the external time-dependent potential $U(\mathbf{r},t).$ The
particle current $\mathbf{j}$ \cite{particle-current} and
particle-conservation equation are given by
\begin{align}
&  \mathbf{j}=-D\left(
\mathbf{\nabla}n+\frac{\mathbf{\nabla}U}{T}n\right)
,\\
&  \frac{\partial n}{\partial t}-D\mathbf{\nabla}\left(
\mathbf{\nabla }n+\frac{\mathbf{\nabla}U}{T}n\right)  =0,
\end{align}
where $D$ is the diffusion constant.

In the absence of the moving potential the particle current is
naturally equal to zero and the equilibrium density
$n_{0}(x)\propto\exp(-U_{0}(x)/T)$.\cite{T-unit} Introducing the
perturbation for the density $\nu(\mathbf{r},t)$ and current due to
the moving potential
\begin{subequations}
\begin{align}
n(\mathbf{r},t)  &  =n_{0}(x)\left[  1+\nu(\mathbf{r},t)\right]  ,\\
\mathbf{\tilde{j}}  &  =-D\left(
\mathbf{\nabla}\nu+\frac{\mathbf{\nabla }\tilde{U}}{T}\right)
n_{0},
\end{align}
we have the following equation for the small perturbation
\end{subequations}
\begin{equation}
\frac{\partial\nu}{\partial t}n_{0}-D\mathbf{\nabla}\left(  n_{0}
\mathbf{\nabla}\nu\right)  =\frac{D}{T}\mathbf{\nabla}\left(  n_{0}
\mathbf{\nabla}\tilde{U}\right)  .
\end{equation}
Using the complex presentation
\begin{align*}
\tilde{U}(\mathbf{r},t)&=\tilde{U} (x)\operatorname{Re}\left\{
\exp\left[  i(ky-\omega t)\right] \right\},\\
\nu(\mathbf{r},t)&=\operatorname{Re}\left\{
\nu_{\omega}(x)\exp\left[ i(ky-\omega t)\right]  \right\},
\end{align*}
we obtain equation for the complex amplitude
$\nu_{\omega}(x)$
\begin{align}
&\left(  -i\eta\omega+Tk^{2}\right)
\nu_{\omega}-\frac{T}{n_{0}}\nabla _{x}\left(
n_{0}\nabla_{x}\nu_{\omega}\right) \notag\\
&=\frac{1}{n_{0}}\nabla _{x}\left(  n_{0}\nabla_{x}\tilde{U}\right)
-k^{2}\tilde{U} .\label{SmallDenCorrection}
\end{align}
where $\eta=T/D$ is the viscosity coefficient.

Assuming that the frequency is small, we solve this equation by
iterations. The first iteration corresponds to quasiequilibrium
\[
\nu_{\omega}^{(0)}=-\frac{\tilde{U}}{T},\ \mathbf{j}^{(0)}=0.
\]
The next iteration has to be found from equation
\begin{equation}
D\left(  \frac{1}{n_{0}}\nabla_{x}\left(
n_{0}\nabla_{x}\nu_{\omega} ^{(1)}\right)
-k^{2}\nu_{\omega}^{(1)}\right)  =i\omega\frac{\tilde{U}}
{T}.\label{FirstIterEq}
\end{equation}
Near the minimum of the potential the equilibrium density can be
represented as
\begin{align*}
n_{0}(x)  &  =n_{00}\exp\left[  -\frac{Kx^{2}}{2T}\right]  ,\\
n_{00}  &  =n_{\xi}\exp\left(  -U_{\min}/T\right)  .
\end{align*}
We also assume that the typical scale of $\tilde{U}(x)$ variation is
larger than confining length of the static potential $\sqrt{T/K}$ so
that we can use the expansion near the minimum,
$\tilde{U}(x)\approx\tilde{U}_{0}-\tilde{F}x$. In this case Eq.
(\ref{FirstIterEq}) has simple analytical solution
\begin{equation}
\nu_{\omega}^{(1)}=-\frac{i\omega\tilde{U}_{0}}{Dk^{2}T}+\frac{i\omega
\tilde{F}x}{D\left(  K+Tk^{2}\right)  },\label{OscDensAmpl}
\end{equation}
which gives the oscillating particle-current amplitudes
\begin{subequations}
\begin{align}
\tilde{j}_{x}  &  =-\frac{i\omega\tilde{F}}{K+Tk^{2}}n_{0},\label{OscCurr_x}\\
\tilde{j}_{y}  &
=ik\frac{i\omega\tilde{U}}{Tk^{2}}n_{0}.\label{OscCurr_y}
\end{align}
The energy dissipation in steady state can be evaluated as
\end{subequations}
\begin{equation}
W\!=\!-\!\int dx\left\langle \mathbf{j\nabla}U\right\rangle
\!=\!\eta\int dx\left\langle \frac{\mathbf{j}^{2}}{n}\right\rangle
\!\approx\!\frac{\eta}{2}\int dx\frac
{|j_{x}|^{2}\!+\!|j_{y}|^{2}}{n_{0}}.
\end{equation}
Substituting currents (\ref{OscCurr_x}) and (\ref{OscCurr_y}), we
finally obtain
\begin{align}
&W =\frac{\eta\omega^{2}}{2}\left(  \frac{\tilde{F}^{2}}{\left(
K+Tk^{2}\right)  ^{2}}+\frac{\tilde{U}^{2}}{T^{2}k^{2}}\right)  \int
dxn_{0}(x)\nonumber\\
&  \approx\frac{\eta\omega^{2}}{2}n_{\xi}\exp\left(
\!-\!\frac{U_{\min}} {T}\right)  \sqrt{\frac{2\pi T}{K}}\left(
\frac{\tilde{F}^{2}}{\left( K\!+\!Tk^{2}\right)
^{2}}\!+\!\frac{\tilde{U}^{2}}{T^{2}k^{2}}\right) .\label{EnDissip}
\end{align}
This formula determines energy losses caused by arbitrary moving
periodic potential near the minimum of the static potential. We will
use it to evaluate the contribution of the fluctuating PVs to the JV
lattice conductivity. For this, in the following sections, we will
obtain explicit expressions for the static and dynamic potentials.

\subsection{Energy of pancake vortex \label{Sec:PVEnergy}}

\subsubsection{Energy profile of pancake vortex near surface in
large-size samples\label{Sec:PVEnergy} }

If the perpendicular magnetic field $H_{z}$ is slightly lower than
$H_{c1} ^{c}$, the PV stacks (Abrikosov vortices) do not penetrate
into the sample, but the PVs can exist near the surface at the
fluctuation regime. The energy of the pancake vortex at the distance
$x$ from the surface is given by \cite{BuzFeinb92}
\begin{align}
&U_{0}(x) \approx\frac{s\Phi_{0}^{2}}{\left(  4\pi\lambda\right)
^{2}} \ln\frac{x}{\xi}-\frac{s\Phi_{0}H_{z}}{4\pi}\left[
1-\exp\left(  -\frac
{x}{\lambda}\right)  \right] \nonumber\\
&  =\frac{s\Phi_{0}\left(  H_{c1}^{c}\!-\!H_{z}\right)
}{4\pi}\!+\!\frac{s\Phi _{0}^{2}}{\left(  4\pi\lambda\right)
^{2}}\ln\frac{x}{\lambda}\!+\!\frac {s\Phi_{0}H_{z}}{4\pi}\exp\left(
-\frac{x}{\lambda}\right)  ,\label{PancEner}
\end{align}
where $s$ is the distance between superconducting layers and
$H_{c1}^{c}
\approx[\Phi_{0}/(4\pi\lambda^{2})][\ln(\lambda/\xi)+0.5]$ is the
lower critical field. Above certain field this potential will have a
minimum at distance $\sim\lambda$ from the surface, see schematic
Fig.\ \ref{Fig-JVPVSurfSchem}(d). Taking derivative,
\[
\frac{dU_{0}}{dx}=\frac{s\Phi_{0}}{4\pi\lambda}\left[
\frac{\Phi_{0}} {4\pi\lambda^{2}}\frac{\lambda}{x}-H_{z}\exp\left(
-\frac{x}{\lambda}\right) \right]  ,
\]
we find the condition for the minimum
\[
u\exp(-u)=\frac{\Phi_{0}}{4\pi\lambda^{2}H_{z}},
\]
with $u=x/\lambda.$ The minimum exists if
\begin{equation}
H_{z}>H_{s}=\frac{e\Phi_{0}}{4\pi\lambda^{2}}.\label{MinField}
\end{equation}
Above this field the pancake energy at the minimum is determined by
equations
\begin{align}
&  U_{\min}= \frac{s\Phi_{0}}{4\pi}\left(  H_{c1}^{c}-H_{z}\right)
+ \varepsilon_{p} \left[  \ln\left(  u_{\min}\right)
+\frac{1}{u_{\min}
}\right]  ,\label{Umin}\\
&
u_{\min}\exp(-u_{\min})=\frac{\Phi_{0}}{4\pi\lambda^{2}H_{z}}.\nonumber
\end{align}
where $\varepsilon_{p}=s\Phi_{0}^{2}/\left(  4\pi\lambda\right)
^{2}$ is the PV energy scale. As $u_{\min}$ roughly behaves as
$\ln(4\pi\lambda^{2} H_{z}/\Phi_{0})$, the main dependence of
$U_{\min}$ on $H_{z}$ is given by the first term, i.e., the minimum
energy approximately linearly decreases with the field.
For Bi2212 $s\approx 1.56$ nm and we estimate the slope of this
dependence as $dU_{\min}/dH_{z}\approx -18.6\ $K/G.

Expanding the potential (\ref{PancEner}) near the minimum,
$U(x)=U_{\min }+K\left(  x-x_{\min}\right)  ^{2}/2$, we evaluate the
spring constant $K$ as
\begin{align}
K  &  =\frac{d^{2}U}{dx^{2}}=\frac{\varepsilon_{p}}{\lambda^{2}}
a_{K},\label{spring const}\\
a_{K}  &
=\frac{1}{u_{\min}}-\frac{1}{u_{\min}^{2}}\approx\frac{1}{\ln
h}.\nonumber
\end{align}
with $h=4\pi\lambda^{2}H/\Phi_{0}$.

The condition of the vortex line formation is roughly $U_{\min}=0$.
In such a case the pancakes will accumulate near the surface and
form a vortex line. The magnetic field $h_{1}$ at which this happens
is determined by equations
\begin{align*}
u_{1} &  =\ln h_{1}-\ln u_{1},\\
h_{1} &  =\frac{\ln\left(  \kappa u_{1}\right)  }{1-\exp\left(
-u_{1}\right) },
\end{align*}
which can be solved by iterations
\begin{align*}
u_{1} &  \approx\ln h_{1}-\ln\ln h_{1},\\
h_{1} &  \approx\ln\kappa+\ln\left[  \ln\left(  \ln\kappa\right)
\right]  .
\end{align*}
Therefore, this field is only slightly larger than $H_{c1}^{c}$. The
Bean-Livingston barrier in layered superconductors with high
anisotropy ratio is eliminated at sufficiently high
temperatures.\cite{BSCCOSurfBarr} In the field interval
$H_{s}<H_{z}<H_{c1}^{c}$ we may expect the accumulation of the
fluctuating PVs near the surface and their density would increase
dramatically when $H_{z}$ approaches $H_{c1}^{c}.$

\subsubsection{Energy of pancake vortex in small-size mesas \label{Sec:FinSize}}

The previous analysis considered the case of the samples much larger
than the London penetration depth $\lambda$. When their dimensions
are comparable with $\lambda$ (which may be the case for the small
mesas) the minima of the PVs energy near the surfaces merge and the
dependence (\ref{Umin}) change. We may illustrate such crossover in
the case of the strip of the width $w$. Using the results of the
paper \cite{Kogan} where the vortex energy in the narrow
superconducting strip was calculated, we write the expression for
the energy of the pancake at the distance $x$ from the strip edge as
\begin{align}
U_{0}(x)&=\frac{s\Phi_{0}\left(  H_{c1}^{c}-H_{z}\right)
}{4\pi}+\frac {s\Phi_{0}^{2}}{\left(  4\pi\lambda\right)
^{2}}\ln\left[  \frac{w} {\pi\lambda}\sin\left(  \frac{\pi
x}{w}\right)  \right]  \notag \\
&+\frac{s\Phi_{0} H_{z}}{4\pi}\frac{\cosh\left(
\frac{x-w/2}{\lambda}\right) }{\cosh\left( \frac{w}{2\lambda}\right)
}. \label{PancEnerSrip}
\end{align}
In contrast to wide strips which have two energy minima near the
edges, the energy of narrow strip at high fields has only one
minimum at the strip center. Two minima merge in the center when the
magnetic field reaches the typical value $H_{s0}$,
\begin{equation}
H_{s0}=\frac{\pi\Phi_{0}}{4w^{2}}\cosh\left(
\frac{w}{2\lambda}\right), \label{HMinCenter}
\end{equation}
and above this field the minimum energy is given by its value at
$x=w/2$,
\begin{align}
U_{\min}&=\frac{s\Phi_{0}}{4\pi}\left[  H_{c1}^{c}-H_{z}\left(
1-\frac {1}{\cosh(w/2\lambda)  }\right)  \right]  \notag \\
&+\frac {s\Phi_{0}^{2}}{\left(  4\pi\lambda\right)  ^{2}}\ln\left(
\frac{w} {\pi\lambda}\right).\label{UminStrip}
\end{align}
At the field
\begin{equation}
H_{s1}\approx \frac{\Phi_{0}}{  4\pi\lambda ^{2}}
\left[\ln\left(\frac{ w }{\pi \xi}\right)+0.5\right]
\frac{\cosh(w/2\lambda)}{\cosh(w/2\lambda)-1} \label{Hs1}
\end{equation}
$U_{\min}$ reaches zero corresponding to condition of fast stack
formation in the center. This field is slightly larger than the
lower critical field for the strip, $H_{c1}^{st}$, which is
determined by the energy of the complete PV stack at the center.  In
particular, for narrow strips $\left(  w<\lambda\right)$
$H_{s1}\approx H_{c1}^{st}\approx [2\Phi_{0}/(\pi w^{2})]
\{\ln\left[ w/(\pi \xi) \right]+0.5\}$, see, e.g., Ref.\
\onlinecite{Abrikos}. The formation of the PV stacks starts at the
center when the field $H_{s0}$ becomes smaller than $H_{c1}^{st}$
giving the approximate condition
\[
\left[\cosh\left(\frac{w}{2\lambda}\right)-1\right]
\frac{\pi^2\lambda^2}{w^2}< \ln \kappa  + 0.5.
\]
For $\kappa\approx100$ this happens already for rather wide strips,
$w\lesssim9\lambda$. Note that at this value  the finite-size
correction to the penetration field is small, $H_{c1}^{st}\approx
H_{c1}^{c}$.

Comparing the field dependences (\ref{Umin}) and (\ref{UminStrip}),
we see that in the narrow strips the slope of the $U_{\min }$ vs
$H_{z}$ dependence should be smaller by the geometrical factor
$\left[ 1-1/\cosh\left(  w/2\lambda\right)  \right]  <1$. Due to the
temperature dependence of $\lambda$, we may expect to observe the
crossover between these regimes for the strips with
$w\gtrsim\lambda(0)$ by varying the temperature. Note also that when
the minimum $U_{0}(x)$ just appears at the center of the strip at
$H_{z}=H_{s0}$, the spring constant $K$ vanishes.

For mesas small in both x and y directions the geometrical factor
will be different, but a qualitative trend is expected to be the
same. To illustrate this case, we consider the cylindrical geometry
which also can be treated analytically. The energy of the pancake
vortex at distance $\rho$ from the center of mesa with radius $R$
may be calculated in the same way as the energy of the vortex in
small superconducting disk, using an image method \cite{BuzBrison}.
In the result we have
\begin{align}
U_{0}(\rho)&=\frac{s\Phi_{0}\left(  H_{c1}^{c}-H_{z}\right)  }{4\pi}
+\frac{s\Phi_{0}^{2}}{\left(  4\pi\lambda\right)  ^{2}}\ln\left(
\frac {R^{2}-\rho^{2}}{R\lambda}\right) \notag \\
&+\frac{s\Phi_{0}H_{z}}{4\pi}\frac {I_{0}\left(  \rho/\lambda\right)
}{I_{0}\left(  R/\lambda\right)}, \label{CylMesaEn}
\end{align}
where $I_{0}(z)$ is the modified Bessel function. The minimum energy
for the fluctuating pancakes is shifted to the center of the mesa if
the field is sufficiently large. With the help of expression
(\ref{CylMesaEn}), we find that it occurs at fields
\begin{equation}
H_{z}>H_{s0}^{\mathrm{cyl}}=\frac{\Phi_{0}}{\pi R^{2}} I_{0} \left(
\frac {R}{\lambda}\right)  . \label{Hs0cyl}
\end{equation}
The lower critical field of the cylinder $H_{c1}^{\mathrm{cyl}}$ was
calculated in Ref.\ \onlinecite{Bobel}
\[
H_{c1}^{\mathrm{cyl}}=\frac{I_{0}\left(  R/\lambda\right)
}{I_{0}\left( R/\lambda\right)
-1}\frac{\Phi_{0}}{4\pi\lambda^{2}}\left[  \ln \kappa
+0.5-\frac{K_{0}\left(  R/\lambda\right)  } {I_{0}\left(
R/\lambda\right)  }\right]  .
\]
In the limit $R\ll$ $\lambda$ the lower critical field is $H_{c1}
^{\mathrm{cyl}}=[\Phi_{0}/(\pi R^{2})][\ln\left( R/\xi\right) +0.38]
$ and the regime of the accumulation of the fluctuating pancakes at
the center of mesa is always realized when the field $H_{z}$ starts
to approach to $H_{c1} ^{\mathrm{cyl}}$. For $\kappa\approx100$ the
condition $H_{s0}^{\mathrm{cyl} }<H_{c1}^{\mathrm{cyl}}$ corresponds
to the mesas with $R\lesssim5.3\lambda$. Note that numerical value
of the critical diameter $2R\approx 10.6\lambda$ is rather close to
the critical width $w\approx 9\lambda$ for the strip geometry. For
smaller mesas at fields just below $H_{c1}^{\mathrm{cyl}}$ the
minimum energy is realized for the PV located in the center,
\begin{align}
U_{\min}&=U_{0}(0)=\frac{s\Phi_{0}H_{c1}^{c}}{4\pi}+\frac{s\Phi_{0}^{2}}{\left(
4\pi\lambda\right)  ^{2}}\ln\left(  \frac{R}{\lambda}\right) \notag
\\
&-\frac{s\Phi _{0}H_{z}}{4\pi}\left[  1-\frac{1}{I_{0}\left(
R/\lambda\right)  }\right]  ,
\end{align}
meaning that the slope of the $U_{\min}$ vs $H_{z}$ dependence in
this regime should be smaller by the factor $\left[
1-1/I_{0}(R/\lambda)  \right]  $, comparing to the large mesas. This
slope again may substantially vary with temperature due to the
temperature dependence of $\lambda$.  Overall behavior is similar to
the stripe case. For the cylindrical mesa with the same diameter as
the stripe width the slope reduction is close and somewhat smaller
than one for the stripe.

\subsection{Interaction of a pancake vortex near surface with dense Josephson
vortex lattice\label{Sec:PV-JV}}

We consider the case of the dense JV lattice when the JVs cores
overlap. This situation realizes when the in-plane magnetic field
$B_{\parallel}$ is larger than the characteristic field
$B_{cr}=\Phi_{0}/\left(  2\pi\gamma s^{2}\right)  $, where the
anisotropy ratio $\gamma=\lambda_{c}/\lambda$ determines the
Josephson length $\lambda_{J}$ $=\gamma s$.\cite{BulClem91} The
interaction of dense JV lattice with PV-stack lattice was first
considered in Ref.\ \onlinecite{BulaevskiiMaley}. The phase
distribution for the triangular Josephson vortex lattice at this
high-field regime in presence of pancake vortices is determined by
equations
\begin{align}
&  \lambda_{J}^{2}\Delta\phi_{n}\!+ \sin\left(
\phi_{n+1}\!-\!\phi_{n}\!+\!\phi
_{v,n+1}\!-\!\phi_{v,n}\!-\!k_{H}(y\!-\!v)\right) \nonumber\\
&  +\sin\left(
\phi_{n}\!-\!\phi_{n-1}\!+\!\phi_{v,n}\!-\!\phi_{v,n-1}\!-\!k_{H}(y\!-\!v)\right)
\!=\!0\label{JVphaseEq}
\end{align}
where $k_{H}=2\pi sB_{\parallel}/\Phi_{0}$ and $\phi_{v,n}$ are the
pancake-vortex phases. The parameter $v$ describes displacement of
the JV lattice. In these equations the $\pi$ phase shift between the
phase differences in the neighboring junctions is already taken into
account.

We consider a single pancake in the layer $n=0$ in a large-size
sample at distance $x_{0}$ from the surface
\[
\phi_{v,n}=\delta_{n,0}\left(  \arctan\frac{y}{x-x_{0}}-\arctan\frac
{y}{x+x_{0}}\right)  ,
\]
where the second term is the contribution from the mirror image,
which is required to satisfy the condition of vanishing normal
current at the surface $\partial \phi_{v,n}/\partial x=0$ for $x=0$.
At high in-plane fields the Josephson current can be treated as a
small perturbation and Eq. (\ref{JVphaseEq}) can be solved by
iterations. The first iteration for the $0$-th layer obeys the
equation
\begin{equation}
\lambda_{J}^{2}\Delta\phi_{0}^{(1)}=2\cos\phi_{v,0}\sin\left[  k_{H}
(y-v)\right] ,\label{eq for phase}
\end{equation}
where
\[
\cos\left[  \phi_{v,0}(\mathbf{r})\right]  =\frac{r^{2}-x_{0}^{2}}
{\sqrt{\left(  r^{2}+x_{0}^{2}\right)  ^{2}-4x^{2}x_{0}^{2}}}.
\]
Without PV the first-order phase iteration is given by
$\phi_{0}^{(1)}(\mathbf{r})\!=\!-2 (\lambda_{J}k_{H})^{-2}\sin\left[
k_{H}(y\!-\!v)\right]  $. The solution of Eq.\ (\ref{eq for phase})
can be written as
\begin{align}
&
\phi_{0}^{(1)}(\mathbf{r})=-\frac{2}{\lambda_{J}^{2}k_{H}^{2}}\sin\left[
k_{H}\left(  y-v\right)  \right] \nonumber\\
& +\frac{1}{\pi\lambda_{J}^{2}}\int d\mathbf{r}^{\prime}\ln\frac
{|\mathbf{r}-\mathbf{r}^{\prime}|}{r_{0}}\left(  \cos\left[
\phi_{v,0}\left( \mathbf{r}^{\prime}\right)  \right]  \!-\!1\right)
\sin\left[  k_{H}(y^{\prime }\!-v)\right]
\end{align}
where the integration is in the limits $-\infty<
x^{\prime},y^{\prime}<\infty$ and the region $x^{\prime}<0$ accounts
for the image contribution. The force acting on pancake is
determined by the phase gradient at its position,
\begin{equation}
f_{\alpha}=\frac{2s\Phi_{0}^{2}}{\left(  4\pi\lambda\right)  ^{2}}
e_{\alpha\beta z}\nabla_{\beta}\phi_{0}^{(1)}\label{PancForce}
\end{equation}
where $e_{\alpha \beta \gamma}$ is the Levi-Civita symbol
($e_{xyz}=-e_{yxz}=1$ and $e_{\alpha \alpha z}=0$). Computing the
phase gradient
and taking it at the pancake position $\mathbf{r}=(x_{0},0)$, we
obtain
\begin{subequations}
\begin{align}
& \nabla_{x}\phi_{0}^{(1)}(x_{0},0) =\frac{\sin\left(  k_{H}v\right)
}
{\pi\lambda_{J}^{2}k_{H}}\mathcal{I}_{x}(k_{H}x_{0}),\\
& \nabla_{y}\phi_{0}^{(1)}(x_{0},0)=\frac{\cos\left(  k_{H}v\right)
}
{\pi\lambda_{J}^{2} k_{H}}\mathcal{I}_{y}(k_{H}x_{0}),\label{PhaseGrads}\\
& \mathcal{I}_{x}(x_{0}) =\!-\!\int d\mathbf{r}\frac{x_{0}-x}{\left(
x_{0}\!-\!x\right)
^{2}\!+\!y^{2}}\frac{r^{2}-x_{0}^{2}}{\sqrt{\left(  r^{2}
+x_{0}^{2}\right)  ^{2}\!-4x^{2}x_{0}^{2}}}\cos y,\label{Ix}\\
& \mathcal{I}_{y}(x_{0}) =\int d\mathbf{r}\frac{y}{\left(
x_{0}-x\right) ^{2}+y^{2}}\frac{r^{2}-x_{0}^{2}}{\sqrt{\left(
r^{2}+x_{0}^{2}\right) ^{2}-4x^{2}x_{0}^{2}}}\sin y.\label{Iy}
\end{align}
\end{subequations}

\begin{figure}[ptb]
\begin{center}
\includegraphics[width=3.2in]{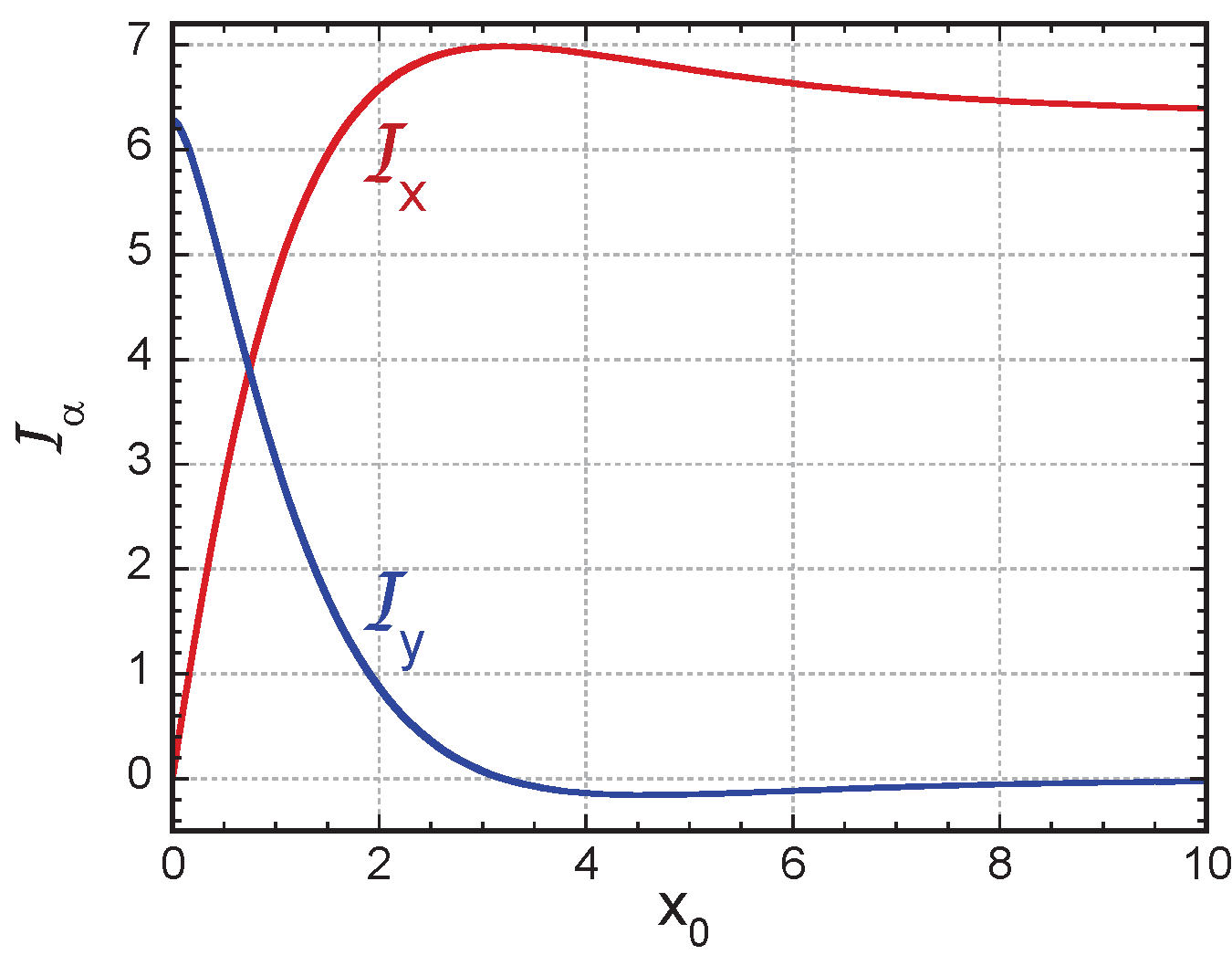}
\end{center}
\caption{(color online) Numerically computed integrals
$\mathcal{I}_{x} (x_{0})$ and
$\mathcal{I}_{y}(x_{0})=\mathcal{I}_{x}^{\prime}(x_{0})$ defined by
Eqs.\ (\ref{Ix}) and (\ref{Iy}), which determine the force acting on
pancake vortex near surface by the dense Josephson vortex lattice. }
\label{Fig-IxIy}
\end{figure}

These integrals have the followings asymptotics: $\mathcal{I}_{y}
(x_{0})\rightarrow2\pi$, $\mathcal{I}_{x}(x_{0})\rightarrow0$ for
$x_{0}\rightarrow0$ and $\mathcal{I}_{y}(x_{0})\rightarrow0$,
$\mathcal{I} _{x}(x_{0})\rightarrow2\pi$ for
$x_{0}\rightarrow\infty$. Moreover, as the force components have to
be derivatives of the same potential, one can demonstrate that the
integrals are connected as $\mathcal{I}_{y}
(x)=\mathcal{I}_{x}^{\prime}(x). $ Numerically computed integrals
are plotted in Fig.\ \ref{Fig-IxIy}.

Finally, the main result of this section is the expression for the
potential energy of the PV in presence of JV lattice (compare with
Eq.\ (\ref{Potential})):
\begin{equation}
\widetilde{U}(\mathbf{r})=\frac{2s\Phi_{0}^{2}}{\left(
4\pi\lambda\right) ^{2}}
\frac{\mathcal{I}_{x}(k_{H}x)}{\pi\lambda_{J}^{2}k_{H}^{2}}\cos\left(
k_{H}y\right)  .\label{PvJvPotEnergy}
\end{equation}
The interaction between the JV lattice and PV near the minimum of
the surface potential is determined by the reduced in-plane field
\begin{equation}
h_{\parallel}\equiv k_{H}\lambda=2\pi s\lambda B_{\parallel}/\Phi
_{0}.\label{reducedHpar}
\end{equation}
Taking typical values $\lambda\approx0.3\mu$m at 70 K and
$B_{\parallel}=1$T we estimate $h_{\parallel}\approx1.42$.

\subsection{Contribution of fluctuating pancakes vortices near surface to the
conductivity of the Josephson vortex lattice \label{Sec:JVConduct}}

Interaction with the surface pancake vortices enhances dissipation
of the Josephson vortex lattice. Additional channel of dissipation
gives additional contribution to the JV conductivity, $\sigma_{JV}=$
$\sigma_{JV}^{(0)} +\sigma_{p}$, where $\sigma_{JV}^{(0)}$ is the JV
conductivity in the absence of PVs and $\sigma_{p}$ is the excess
conductivity due to the interaction with pancakes. The corresponding
pancake surface contribution, $\sigma_{p}$, can be evaluated using
energy-dissipation formula (\ref{EnDissip})
\begin{align}
W&=\sigma_{p}E_{z}^{2}L_{x}L_{y}s
=\!2\frac{\eta\omega^{2}}{2}n_{\xi}L_{y} \exp\left(
\!-\!\frac{U_{\min}}{T}\right)\!  \sqrt{\frac{2\pi T}{K}}\notag \\
&\times\left( \frac{\tilde{F}^{2}}{\left(  K\!+\!Tk_{H}^{2}\right)
^{2}}\!+\!\frac{\tilde{U}^{2} }{T^{2}k_{H}^{2}}\right)  .
\end{align}
Here $W$ is the energy dissipated per single superconducting layer,
$L_{x}$ and $L_{y}$ are the dimensions of crystal along the
corresponding axis and the coefficient 2 comes from the contribution
from two $yz$ surfaces.

Taking in mind the relation between electric field and velocity for
the moving JV lattice $E_{z}=vH_{x}/c=(\Phi_{0}/2\pi cs)\omega$ and
extracting parameters $\tilde{F}$ and from Eq.\
(\ref{PvJvPotEnergy}), we obtain for the excess conductivity
\begin{align}
\sigma_{p}=&\left(  \frac{2\pi cs}{\Phi_{0}}\right) ^{2}\frac{\eta
n_{\xi}}{L_{x}s}\!\exp\left( - \frac{U_{\min}}{T}\right) \notag
\\
&\times\sqrt{\frac{2\pi T}
{a_{K}\varepsilon_{p}}}\frac{4\lambda^{7}}{\pi^{2}\lambda_{J}^{4}
h_{\parallel}^{2}}G\left[T/\varepsilon_{p},h_{\parallel}
\right],\label{excess conductivity}
\end{align}
where
\[
G[\tilde{T},h_{\parallel}]=\frac{\left[  \mathcal{I}_{x}^{\prime}
(h_{\parallel} u_{\min})\right]  ^{2}}{\left(
a_{K}+\tilde{T}h_{\parallel }^{2}\right)  ^{2}
}+\frac{\mathcal{I}_{x}^{2}(h_{\parallel} u_{\min})}
{\tilde{T}^{2}h_{\parallel}^{4}},
\]
$u_{\min}\approx\ln h$, and $\tilde{T}=T/\varepsilon_{p}$ is the
dimensionless temperature. The pancake viscosity, $\eta$, also
determines the the flux-flow resistivity for the in-plane current,
$\rho_{ff}=s\Phi_{0}B/(c^{2}\eta)$. A natural scale of conductivity
is the pancake flux-flow conductivity at field
$B=\Phi_{0}/(4\pi\lambda^{2})$,
\[
\sigma_{ff,\lambda}=\frac{4\pi\lambda^{2}c^{2}\eta}{s\Phi_{0}^{2}}.
\]
Using this scale, we can rewrite Eq.\ (\ref{excess conductivity}) in
somewhat more transparent form
\begin{align}
\sigma_{p}=&\sigma_{ff,\lambda}\frac{\lambda}{\pi
L_{x}}s^{2}n_{\xi}\exp\left( -\frac{U_{\min}}{T}\right) \notag \\
&\times \sqrt{\frac{2\pi T}{a_{K}\varepsilon_{p}}}
\frac{\lambda^{4}}{\lambda_{J}^{4}h_{\parallel}^{2}}G[T/\varepsilon
_{p},h_{\parallel}]\label{ExcCond-ff}
\end{align}

In spite of the rather complicated general $\sigma_{p}(T,
H_{x},H_{z})$ dependence (\ref{excess conductivity}), the main
variation of $\sigma_{p}$ just below $H_{c1}^{c}$ is related with
the change of the concentration of the fluctuating pancakes.  In
samples with the lateral sizes larger than $\lambda$ this leads to
the very characteristic exponential $\sigma_{p}$ dependence on $H_z$
\begin{equation}
\sigma_{p} \approx A(H_{x},T)\exp\left( \frac{s\Phi_{0}H_z}{4\pi T}
\right) ,\label{ExpDepend}
\end{equation}
and the dominating temperature dependence of the prefactor is given
by
\[
A(H_{x},T)=\tilde{A}(H_{x},T)\exp\left(
-\frac{s\Phi_{0}H_{c1}^{c}}{4\pi T}\right).
\]
These are the easiest qualitative predictions to compare with
experiment. We remind also that in small-size mesas the expression
in the exponent of Eq. (\ref{ExpDepend}) acquires a geometrical
factor depending on mesa size, as described in Sec.
\ref{Sec:FinSize}.

A more complicated issue is the dependence of the excess
conductivity on the in-plane magnetic field. The general formula
(\ref{ExcCond-ff}) significantly simplifies at high in-plane fields
$h_{\parallel}\gg1$. In this regime we obtain
$G[\tilde{T},h_{\parallel}]\approx4\pi^{2}/(\tilde{T}^{2}h_{\parallel
}^{4})$ and
\begin{equation}
\sigma_{p}\!\approx\sigma_{ff,\lambda}\sqrt{\frac{2\pi}{a_{K}}}\frac{4\pi
\lambda}{L_{x}}s^{2}n_{\xi}\exp\left(\!-\frac{U_{\min}}{T}\right)
\left( \frac{\varepsilon_{p}}{T}\right)
^{3/2}\!\frac{\lambda^{4}}{\lambda_{J}
^{4}h_{\parallel}^{6}},\label{ExcCondHighFields}
\end{equation}
i.e., we expect that at high fields the excess conductivity rapidly
decreases with increasing in-plane field $\propto1/H_{x}^{6}$.

We have considered the influence of the fluctuating pancakes on the
dense moving JV lattice only near the $yz$ surfaces, perpendicular
to the in-plane magnetic field. Naturally, the fluctuating PVs also
exist near the $xz$ surfaces and they also will contribute to
dissipation. The JV lattice will displace them mainly along the $x$
axis. Accurate calculations for this case are more complicated
because one has to take into account deformation of the JV lattice
near the surface. However, even without performing the calculation
of this contribution, it is evident that it will be also
proportional to the PVs concentration and therefore the dependence
(\ref{ExpDepend}) will describe it as well. This dependence also
should describe behavior of the c-axis excess conductivity for
smaller in-plane fields, in the regime of dilute JV lattice. In this
regime the fluctuating PVs enhance dissipation of individual JVs
meaning that $\sigma_p\propto 1/H_x$.

\section{Experiment \label{Sec:Exper}}

\subsection{Experimental setup}

Bi2212 single crystals grown by the traveling solvent floating zone method were fabricated into mesas sandwiched by two superconducting electrodes of Bi2212 (see Fig.\ \ref{Fig-JVPVSurfSchem}a) with a focused ion beam (FIB) machine SMI2050, SII Nanotechnology Inc. The fabrication details are described in Ref.\ \onlinecite{Rhodes-Kakeya}. The studied samples listed in Table \ref{tab1} include two medium-size mesas with lateral sizes $\gtrsim 5$ $\mu$m and four small mesas with lateral sizes $\sim 1$ $\mu$m.

Resistance of the mesa was measured with the four-probe method with either dc or ac current excitation. For the ac excitation, the lock-in detection was used for voltage measurements. External magnetic fields $\bm{H}$ were applied by a split-pair superconducting magnet that generate a horizontal magnetic field up to 80 kOe and the angle $\theta$ between the magnetic field and the $ab$-plane ($x$-axis) was varied by rotating the sample probe with a precision rotator. Perpendicular ($c$-axis) component of the magnetic field $H_{z}$ is given by $H\sin\theta$. In all plots presented here, $H_z$ is always swept from negative to positive, meaning that PVs are expelled from and introduced into the mesas in the negative and positive $H_z$ regions, respectively. Therefore asymmetry with respect to sign of $H_z$ in some plots may be caused by hysteretic behavior.

\begin{table}[tb]
\caption{List of samples. $L$, $W$, and $t$ denote length
($\perp\bm{H}$), width ($\parallel\bm{H}$), and thickness
($\parallel c$) of the mesa, respectively, see Figure
\ref{Fig-JVPVSurfSchem}a. $H_{p}$ is the JV flow oscillation period
experimentally observed. Note that $H_p$ is slightly different from
$\Phi_0/sL$ due to the ambiguity of the mesa dimensions. Units for
dimensions, magnetic fields, and temperature are $\mu$m, kOe, and K,
respectively.} \label{tab1}
\begin{ruledtabular}
\begin{tabular}{lllllll}
& $L$ & $W$ & $t$ &  $H_p$ & $T_c$  \\
\hline
Br0552 & 5.0 & 4.9 & 0.16 & 2.6 & 84.7 \\
Br0573 & 5.0 & 10.2 & 1.0 & 2.58 & 88.4 \\
Br0640 & 1.1 & 1.13 & 0.25  & 11.5 & 84 \\
Br0742 & 1.53 & 1.45 & 0.26 & 8.6 & 86 \\
Br0746 & 0.8 &  2.06 & 0.5 & 17.25 & 86.5 \\
Br0747 & 0.4 &  1.5 & 0.3 & 34.5 & 83.9 \\
\end{tabular}
\end{ruledtabular}
\end{table}

\subsection{Experimental results and discussion}

\begin{figure}[ptb]
\begin{center}
\includegraphics[width=3.4in]{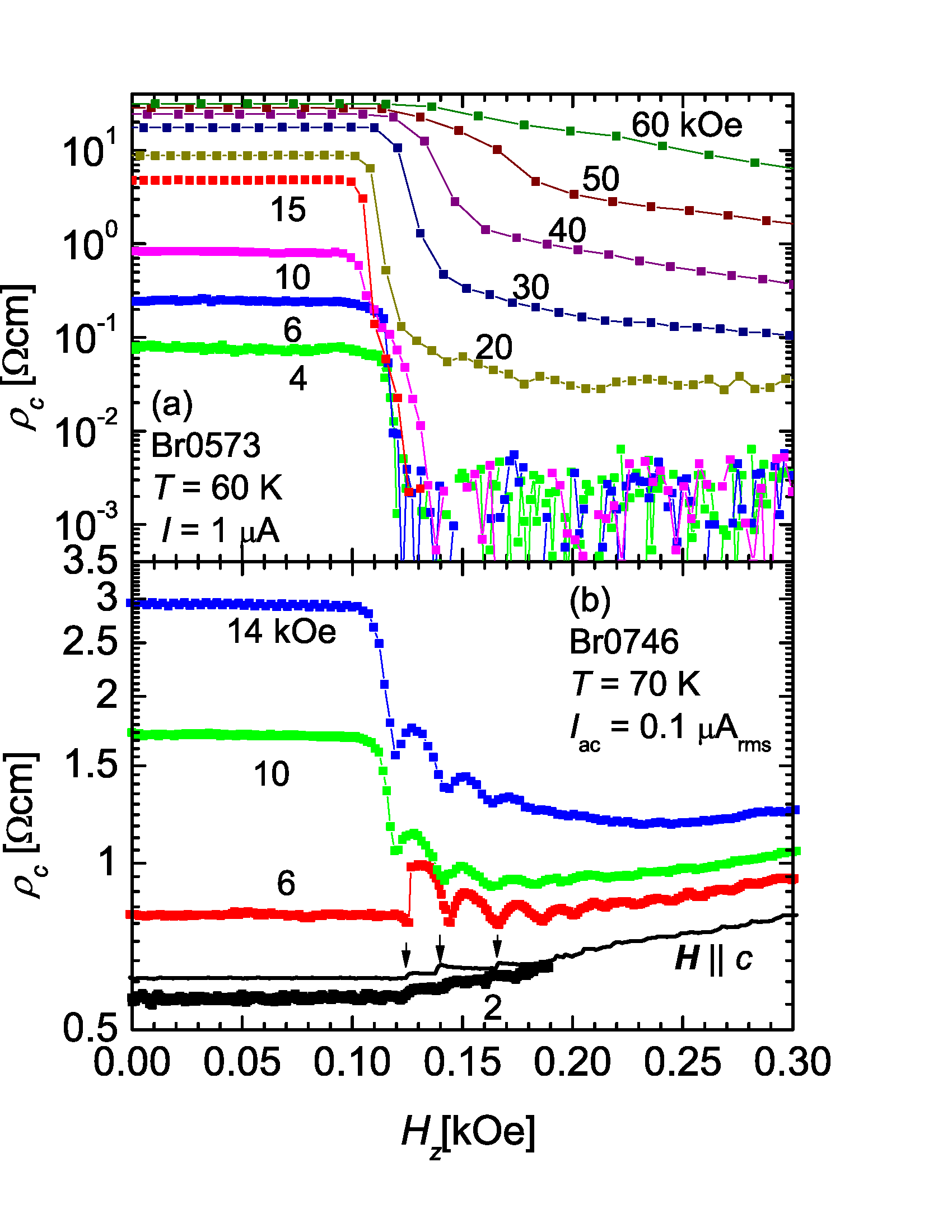}
\end{center}
\caption{Typical dependences of the resistivity $\rho_{c}(H_{z})$ in the
 midsize mesa Br0573 at 60 K (a) and small-size mesa Br0746 at 70 K (b). (a) Sharp drop corresponds to penetration of the PV
stacks. We focus on behavior below this drop. (b) In small-size mesas, in addition to the initial sharp drop, quasi-periodic drops due to penetration of the individual PV stacks are found. In the case of either small magnitude of tilting magnetic field (2 kOe) or magnetic field parallel to the $c$-axis (solid line), stepwise changes in $\rho_c$ indicated by arrows are observed. } \label{0573@60K_rho}
\end{figure}

The $c$-axis resistivity $\rho_{c}$ in the vicinity of $\theta=0$ gives
the JV flow resistance. With increasing $H_{z}$, $\rho_{c}$ suddenly
drops due to penetrating PVs. Without taking fluctuating PVs into
consideration, the value of the field when $\rho_{c}$ drops should
correspond to the lower critical field $H_{c1}$ (corrected by the
demagnetization factor). The demagnetization factor of a small mesas
($L$ or $W\simeq1\mu$m) is small because the thickness of the whole
crystal is about 10 $\mu$m for all samples and it is larger than the
width.
For the midsize mesas  ($L$ or $W\simeq5-10\mu$m), due to
demagnetization effect, the field at the edge of junctions in the
Meissner state should exceed the external field by factor
$\sim$1.2-1.3.
An example of $\rho_{c}$ as a function of $H_{z}$ at different applied
fields for one of the medium-size mesas is plotted in Fig.\
\ref{0573@60K_rho}(a). We see that the steep drop in $\rho_{c}$ in the
low-field region is smeared with increasing parallel field $H$. We
focus here on the quantitative characterization of behavior slightly
below this drop.

Field behavior of small mesas has two important features. Firstly, when the magnetic field is aligned with the layers, due to strong interaction with the edges, the dense JV lattice has rectangular structure almost at all fields and only transforms into the triangular configuration in the vicinity of fields corresponding to integer number of fluxes per junctions, $H_{n}=n H_{p}$ with $H_{p}=\Phi _{0}/sL$.\cite{MachidaPRL06,MagOscTheory} Values of $H_{p}$ for different mesas are listed in the Table \ref{tab1}. The c-axis resistivity has pronounced oscillations as function of $H_{x}$ with maxima located near $H_{n}$.\cite{RectSmallStacks,KakeyaPRB:2009} Secondly, when the c-axis magnetic field is applied, penetration of the individual PV stacks can be resolved. In the case of magnetic field tilting from from the $ab$-plane, a PV stacks penetrating into the mesa strongly impede JV flow along the $ab$-plane, which causes a sharp drops in $\rho_c$. With increasing $H_z$, the separation between the drops decreases and finally corresponds to the period of one flux quantum per mesa area, $\Phi_0/LW$, as shown in Fig.\ref{0573@60K_rho} (b). These quasi-periodic drops in $\rho_c$ are naturally attributed to penetrations of the individual PV stacks into the mesa. Even in the absence of JVs ($\bm{H} \parallel c$), stepwise increase in $\rho_c$ (shown as a solid line) pointed by arrows in Fig. \ref{0573@60K_rho} (b) with similar periodicity was observed. These steps are attributed to additional interlayer phase fluctuation caused by the penetrated PV stacks.\cite{Rhodes-Kakeya} Therefore in small mesas the c-axis resistivity has oscillating dependences on {\em both} field components.

\begin{figure}[ptb]
\begin{center}
\includegraphics[width=3.4in]{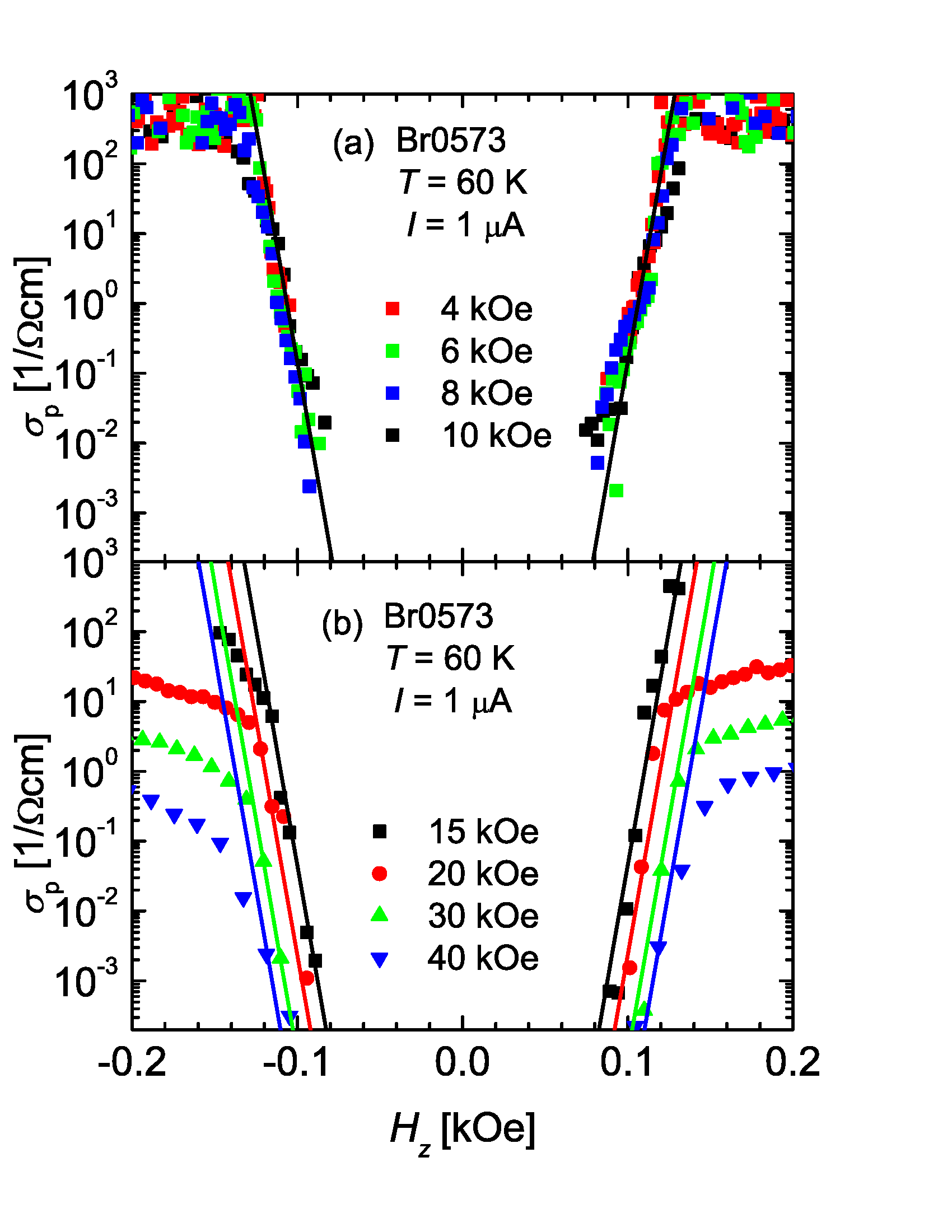}
\end{center}
\caption{The dependences of the excess conductivity on $H_z$ for the
sample  Br0573 at $T=60$ K in (a) $2<H<10$ kOe and (b) $15<H<60$
kOe.
Solid lines describe exponential dependence (\ref{ExpDepend}). One
can see that the data follow the theoretical prediction.
} \label{0573@30kOe}
\end{figure}
In order to quantify the additional dissipation of the JV lattice
due to penetration of fluctuational PVs,
we show the plots of the excess $c $-axis conductivity
$\sigma_{p}(H_{x},H_{z})=\sigma_{c}(H_{x},H_{z})-\sigma_{c}(H_{x},0)$
as a function of the $c$-axis magnetic field $H_{z}$ for several
samples at different temperatures in the following figures.
Here, $\sigma_{c}(H_{x},0)=1/\rho_c(H_{x},0)$ is the conductivity
for the magnetic field  aligned with the layers, at $\theta=0$
($H_{z}=0$).
Only the data at temperatures higher than $T=60$K
are shown because at low temperatures (below 50 K), the surface
barrier and bulk pinning for PVs cause pronounced hysteresis in the
dependences $\rho_{c}(H_{z})$, which makes an analysis quite complicated.

Figure \ref{0573@30kOe} illustrates the evolution of the $H_{z}$
dependence of the excess conductivity $\sigma_p$ with increasing field obtained
from data presented in Fig.\ \ref{0573@60K_rho}. Other mesas show
similar behavior in the penetration region.
In the figure, the predicted exponential dependence
(\ref{ExpDepend}) is plotted as a thick solid line for each $T$. Since the constant
$A$ is the sole fitting parameter, the slope of the semi-log plot of
the experimental data provides a direct test for Eq.\ (\ref{ExpDepend}).
The first general observation is that the experimental data clearly show the existence
of the exponential regime
just before the penetration of the perpendicular field into the mesas.
This exponential increase spans up to five orders of magnitude. The
slopes of the $\ln(\sigma_{p})$  vs $H_{z}$ curves are in good agreement
with Eq.\ (\ref{ExpDepend}). The observation of this regime provides
strong evidence of generation of the fluctuating PV near the sample
surface.

\begin{figure}[ptb]
\begin{center}
\includegraphics[width=3.2in]{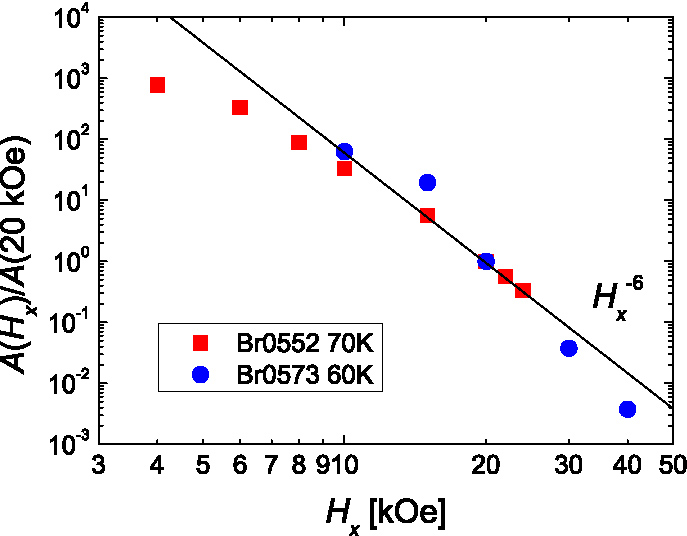}
\end{center}
\caption{
The normalized preexponential factor $A$ in Eq. (\ref{ExpDepend}) as
a function of in-plane field $H_x$ for the mesa Br0552 at 70K and
Br0573 at 60 K. The factor is normalized by its value at $H_x$=20 kOe.
} \label{A_Hx_normalized}
\end{figure}

There are several common features in the experimental data. The
slopes of dependences $\ln\sigma_{p}$ vs $H_{z}$ remain independent
of the in-plane field only up to the typical fields $\sim$40 kOe, in
agreement with theory. At higher fields (not shown) the slope starts
to decrease and at the same time the experimental dependencies
$\sigma_{p}(H_{z})$ are no more symmetric over $\pm H_{z}$. Probably
at such a high magnetic field it is difficult to reliably control
the small $H_{z}$ component. At small fields $\lesssim 10$ kOe the
excess conductivity is only weakly depends on the field while at
higher fields it starts to decrease rapidly. This behavior is most
probably related with the crossover into the dense-lattice regime.
Figure \ref{A_Hx_normalized} shows dependence of the preexponential
factor $A$ on the in-plane field $H_x$ for mesas Br0552 at $T=70$K
and Br0573 at 60 K. One can see that decay at fields $H_x > 6$kOe is
close to theoretical $H_x^{-6}$ dependence given by Eq.
(\ref{ExcCondHighFields}).

\subsubsection{Small mesas: triangular vs rectangular JV lattice}

\begin{figure}[ptb]
\begin{center}
\includegraphics[width=3.4in]{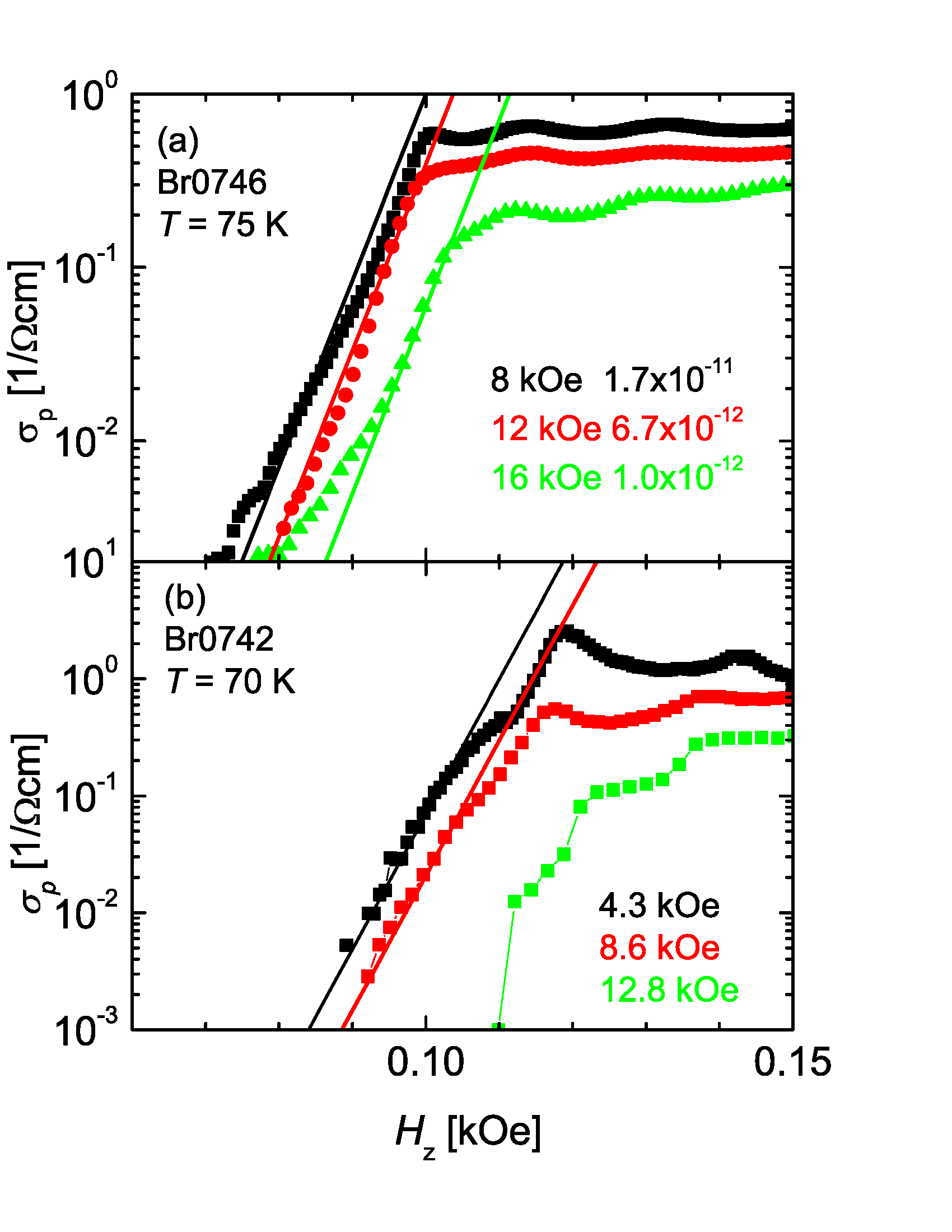}
\end{center}
\caption{The excess pancake conductivity for two small-size mesas.
(a) Data for the mesa Br0746 at 75 K for three values of magnetic
filed smaller than $H_p=17.25$kOe. Theoretical exponential
dependences and corresponding preexponential factors are also shown.
(b) The excess conductivity for the sample Br0742 at 70 K also for
three values of field. For the first two values 4.3 and 8.6 kOe
corresponding to 0.5$H_p$ and 1.0$H_p$ agreement with theory is very
good. There is no agreement for the last value,  12.8 kOe $\approx$
1.5$H_p$, the transition is much sharper and occurs at higher $H_z$.
At this value the rectangular JV lattice is realized in the mesa
which is less sensitive to the fluctuating PVs.}
\label{0746@75K0742@70K}
\end{figure}

An important feature of small-size mesas is the existence of regions
of the rectangular vortex lattice at fields $H_x>B_{\mathrm
cr}L/(\gamma s)$. In this range with increasing magnetic field the
JV lattice undergoes a series of structural phase transitions
between the triangular and rectangular configurations, with
triangular structures located around the fields $nH_p$ and
rectangular structure located around the fields $(n+1/2)H_p$. The
widths of the triangular regions shrink with increasing field
\cite{MachidaPRL06,MagOscTheory}. The triangular configuration is expected to
have much stronger interaction with the PVs than the rectangular
lattice because there are no in-plane currents in the latter state.

Figure \ref{0746@75K0742@70K} presents the $H_z$ dependences of the
excess conductivity for two small-size mesas  Br0746 and  Br0742.
The panel (a) shows data for mesa  Br0746 at  75 K for fields
smaller than $H_p=17.26$ kOe. We can see that there is again a wide
range of exponential dependence and the slope $d\ln \sigma_p/d H_z$
agrees with theoretical predictions. Figure\ \ref{0746@75K0742@70K}(b)
shows the excess conductivity  for the small mesa Br0742 for three
values of field 4.3, 8.6, and 12.8 kOe almost corresponding to 0.5, 1.0,
and 1.5$H_p$. At the last field the rectangular lattice is realized.
We can see that while agrement with the predicted dependence is very
good for the first two values of field, for the last field the
excess conductivity before penetration is much smaller which makes
the transition much sharper. This confirms that the fluctuating
pancakes have much weaker influence on dissipation of the
rectangular vortex lattice.

We finish this subsection with a small comment on the oscillating
behavior of the c-axis resistivity in small-size mesas at larger
$H_z$ due to the penetrating individual PV stacks illustrated in
Fig.\ \ref{0573@60K_rho}(b). This behavior actually appears due to
the competition between two opposite factors. On the one hand,  the
PVs in the stacks directly contribute  to the  dissipation of the JV
lattice  reducing the resistivity. On the other hand, penetration of
the magnetic flux inside the mesa relaxes the Meissner currents near
its surface which reduces the contribution coming from the
fluctuating PVs and increases the resistivity. We can see that
slightly above the penetration field the resistivity increases
meaning that the second mechanism actually prevails in this region.
Moreover, continuous increase of the resistivity for $H=10$ and $14$
kOe indicates that the full PV stack in the mesa center is not
formed at once but continuously builds within finite field range.

\subsubsection{Evolution with increasing temperature and finite-size
effect in small mesas}

\begin{figure}[ptb]
\begin{center}
\includegraphics[width=3.4in]{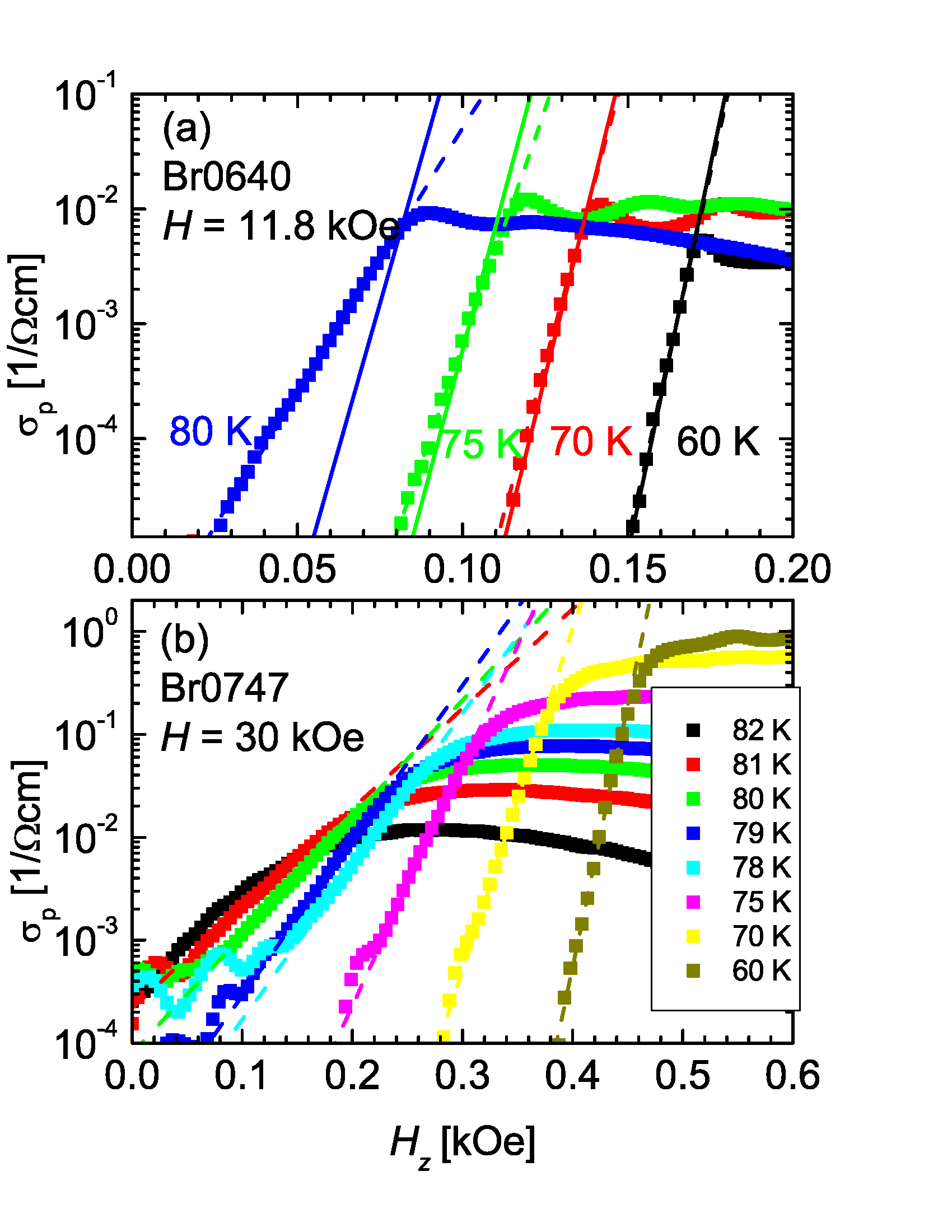}
\end{center}
\caption{ Temperature evolution of the excess conductivity for
small-size mesas. (a) Excess-conductivity plots for the small-size
mesa Br0640 for the field $11.8$ kOe corresponding to the triangular
lattice at four values of temperature, 60, 70, 75, and 80 K. Solid
lines represent exponential dependences given by Eq.\
(\ref{ExpDepend}) at temperatures for corresponding colors, whereas
the broken lines are given by Eq.\ (\ref{ExpDepend_shortjunc}) for
the small-size regime. One can see that the exponential dependencies
agree with the prediction for the large-size regime at 60, 70, and 75 K but
the slope is considerably smaller at 80K. (b) Data for the mesa
Br0747 from 60 to 82 K at 3 T. The exponential dependence
for the small-size  regime (\ref{ExpDepend_shortjunc}) shown by
broken lines describes data very well. }
\label{0640@12kOe0747@30kOe}
\end{figure}

\begin{figure}[ptb]
\begin{center}
\includegraphics[width=3.15in]{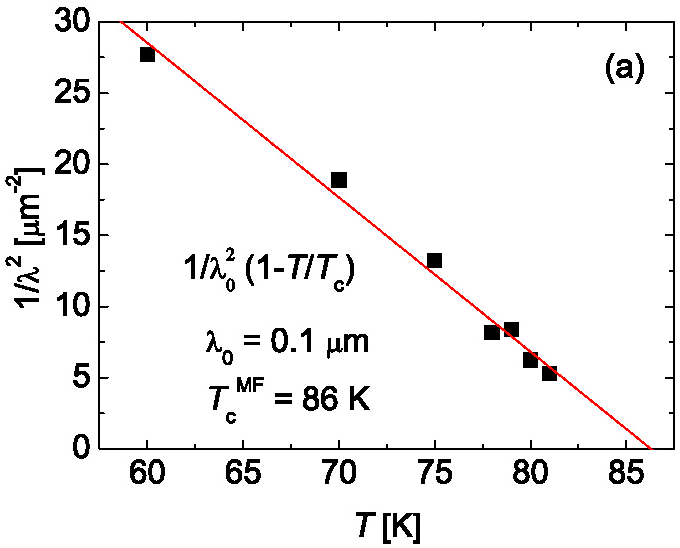}
\includegraphics[width=3.2in]{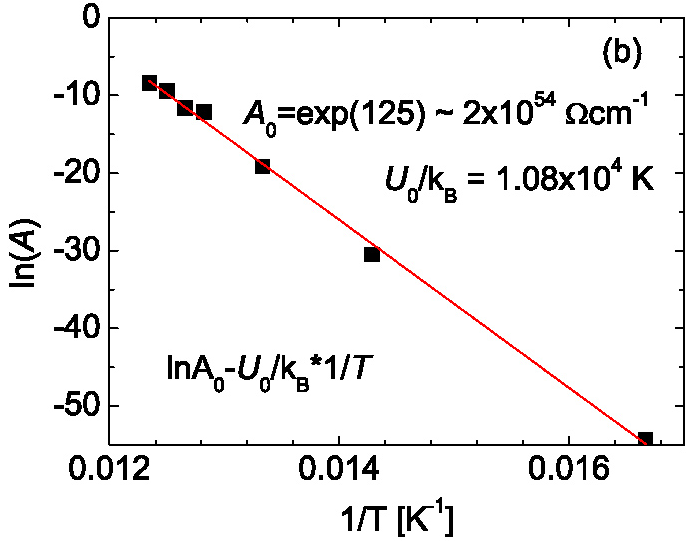}
\end{center}
\caption{ (a) Temperature dependence of $\lambda^{-2}$
extracted from slopes $d\ln{\sigma_p}/dH_z$ for mesa Br0747 using
the activation energy, Eq. (\ref{UminStrip}). (b)
Arrhenius plot of the preexponential factor $A$ for the same
mesa. The origin of huge value of $A_0$ is explained in the text.}
\label{lambdaA0747@30kOe}
\end{figure}
Figure \ref{0640@12kOe0747@30kOe} presents evolution of the
$\sigma_p$ vs $H_z$ dependences with increasing temperature at fixed
field for two small-size mesas. In both cases fields are close to
the corresponding values of $H_p$. We can see that for the mesa
Br0640 (panel (a)) these dependences agree with the prediction for the
large-size regime for temperatures 60, 70, and 75 K  but slope $d
\ln\sigma_p/dH_z$ becomes considerably smaller for 80K. For the
smallest mesa Br0747 with the width $L=0.4$ $\mu$m (panel (b)) the slopes are
smaller than predicted by Eq.\ (\ref{ExpDepend}) for all
temperatures and decrease with increasing temperature. This behavior
can be naturally explained by the finite-size effect considered in
Sec.\ \ref{Sec:FinSize}.
As the other size of this mesa $W$ is 3.75 times larger than the
width $L$, we can apply the strip approximation. In this case when
the London penetration depth $\lambda$ becomes comparable with the
mesa smallest lateral size, the energy minimum of the PV,
$U_{\min}$, is shifted to its center and the slope the $U_{\min}$vs
$H_{z}$ decreases, thus $H_{z}$ must be substituted by $H_{z}\left[
1-\cosh^{-1}\left( L/2\lambda\right)  \right]$:
\begin{equation}
\sigma_{p}\! \approx \! A(H_{x},T)\exp\left\{ \frac{s\Phi_{0}H_{z}}{4\pi T}
\left[ 1\!-\!\cosh^{-1}\!\left( \frac{L}{2\lambda}\right)  \right]
\right\} .\label{ExpDepend_shortjunc}
\end{equation}
This means that from the slopes we can extract the
temperature-dependent $\lambda$. Figure \ref{lambdaA0747@30kOe} (a)
presents the temperature dependence of $\lambda^{-2}$ obtained in
such a way. It shows expected linear dependence near $T_c$,
$\lambda^{-2}=\lambda_0^{-2}(1-T/T_c^{\mathrm{MF}})$, with the
Ginzburg-Landau value $\lambda_0=0.1\mu$m and the mean-field
transition temperature $T_c^{\mathrm{MF}}\approx 86$K which is 2~K
higher than the resistivity transition temperature for this
mesa, $T_c=84$K. Additional information can be obtained from the
temperature dependence of the preexponential factor $A$ presented in
the right part of Fig.\ \ref{lambdaA0747@30kOe}.  As follows from
Eq.\ (\ref{UminStrip}), it has Arrhenius temperature dependence,
$A=A_0 \exp(-U_0/k_BT)$, with the activation energy
$U_0=[s\Phi_0^2/(4\pi\lambda_0)^2][\ln(L/\pi \xi)+0.5]$. This
agrees with experiment and the fit gives
$U_0/k_B=1.08\cdot10^{4}$K. This is consistent with the above
value of $\lambda_0$ if we assume $\xi\approx 50$nm at 75K. Note
also that the huge value of $A_0$ obtained from the fit, $A_0\approx
\exp(125) [\Omega \mathrm{cm}]^{-1}$, is, in fact, very reasonable
because $A_0$ contains the factor
$\exp(U_0/k_BT_c^{\mathrm{MF}})\approx \exp(126)$.
The obtained value of $\lambda_0$ for our overdoped small mesa, however, is somewhat smaller than the values reported in the literature.\cite{AnukoolLonPenPRB09}
Note that the mesa Br0640 also shows the same trend. However for
this mesa the sizes $W$ and $L$ are close so that the narrow strip
model is not applicable. Accurate description of this case requires
numerical analysis.

\section{Conclusions}

Layered superconductors in magnetic fields are characterized by
complex interplay between the Josephson and pancake vortices. In
particular, dissipation of the Josephson vortex lattice is proven to
be extremely sensitive to the presence of pancake vortices. We
utilize this property to probe the fluctuating pancake vortices. In
the Meissner state, in the c-axis magnetic field smaller than the
lower critical field, the pancakes vortices can not form the
Abrikosov vortex lines. In the fluctuational regime, however, the
individual pancake vortices may exist near the surface or inside the
small samples. They lead to additional contribution to the c-axis
conductivity of the Josephson vortex lattice which has very
characteristic exponential dependence on the c-axis magnetic field.
While in mesas with lateral sizes significantly larger than the
London penetration depth $\lambda$ the slope $d\ln(\sigma_p)/dH_z$
of this dependence is universal, in smaller mesas it acquires the
geometrical factor depending on the ratio size/$\lambda$. We
systematically studied the excess c-axis conductivity of the JV
lattice due to the fluctuating pancake vortices in mesas fabricated
out of Bi$_{2}$Sr$_{2}$CaCu$_{2}$O$_{8+\delta}$ crystals. The
predicted exponential dependence is clearly observed and its slope
agrees very well with the theoretical value. Analyzing the
temperature evolution of the slope in the mesa with smallest width
and extracting the geometrical factor, we were able to restore the
temperature dependence of the London penetration depth.
These findings provide strong evidence for the existence of the
fluctuating pancakes in the Meissner state and demonstrate that the
Josephson-vortex lattice provides a unique tool to probe them.

\begin{acknowledgments}
A.E.K. was supported by UChicago Argonne, LLC, operator of Argonne
National Laboratory, a U.S. Department of Energy Office of Science
laboratory, operated under contract No. DE-AC02-06CH11357. A.I.B.
also would like to acknowledge support from Argonne National
Laboratory for the one month visit during which this work was initiated.
 I.K. was supported by Global COE program on photonics
and electronics science and engineering at Kyoto University, Kansai
Research Foundation, and Mazda Foundation.
\end{acknowledgments}


\begin{thebibliography}{99}
\bibitem{demag-footnote}Note that we consider here the superconducting mesas with
low demagnetization factor along $H_{z}$.
\bibitem{particle-current}The particle current $\mathbf{j}$ in this section
describes flow of the pancake vortices proportional to the local
electric field and should not be confused with the electric current.
\bibitem{T-unit} We measure the temperature $T$ in the energy units throughout the
paper.  Conversion to kelvins can be made by usual substituition
$T\rightarrow k_B T$, where $k_B$ is the Boltzmann constant.
\end{thebibliography}
\end{document}